\documentclass[aps,prl,twocolumn,groupedaddress,showpacs,amsmath,amssymb,superscriptaddress,floats]{revtex4}
\usepackage{bm,graphics,amssymb}
\usepackage{graphicx}
\usepackage{dcolumn}
\usepackage{bm}
\usepackage{color}
\def\degree{\ensuremath{^\circ}}

\begin{document}

\preprint{Submitted to Physical Review Letters}
\title{Behavior of colloidal particles at an air/nematic liquid crystal interface}
\author{M.A. Gharbi}
\affiliation{Laboratoire des Collo\"\i des, Verres et Nanomat\'eriaux(LCVN), UMR5587 CNRS and
Universit\'e Montpellier II, Place Eug\`ene Bataillon, 34095 Montpellier, France.}
\author{M. Nobili}\email{nobili@lcvn.univ-montp2.fr}
\affiliation{Laboratoire des Collo\"\i des, Verres et Nanomat\'eriaux(LCVN), UMR5587 CNRS and
Universit\'e Montpellier II, Place Eug\`ene Bataillon, 34095 Montpellier, France.}
\author{M. In}
\affiliation{Laboratoire des Collo\"\i des, Verres et Nanomat\'eriaux(LCVN), UMR5587 CNRS and
Universit\'e Montpellier II, Place Eug\`ene Bataillon, 34095 Montpellier, France.}
\author{G. Pr\'evot}
\affiliation{Laboratoire des Collo\"\i des, Verres et Nanomat\'eriaux(LCVN), UMR5587 CNRS and
Universit\'e Montpellier II, Place Eug\`ene Bataillon, 34095 Montpellier, France.}
\author{P. Galatola}
\affiliation{Laboratoire Mati\`ere et Syst\`emes Complexes (MSC), UMR 7057 CNRS and Universit\'e
Paris Diderot-Paris 7,  CC 7056, 75205 Paris, France.}
\author{JB. Fournier}
\affiliation{Laboratoire Mati\`ere et Syst\`emes Complexes (MSC), UMR 7057 CNRS and Universit\'e
Paris Diderot-Paris 7,  CC 7056, 75205 Paris, France.}
\author{Ch. Blanc} \email{blanc@lcvn.univ-montp2.fr}
\affiliation{Laboratoire des Collo\"\i des, Verres et Nanomat\'eriaux(LCVN), UMR5587 CNRS and
Universit\'e Montpellier II, Place Eug\`ene Bataillon, 34095 Montpellier, France.}
\date{\today}
\pacs{61.30.Jf,61.30.Hn,61.72.Cc}

\begin{abstract}
We examine the behavior of spherical silica particles trapped at an air-nematic liquid crystal
interface. When a strong normal anchoring is imposed, the beads spontaneously form various
structures depending on their area density and the nematic thickness. Using optical tweezers, we
determine the pair potential and explain the formation of these patterns. The energy profile is
discussed in terms of capillary and elastic interactions. Finally, we detail the mechanisms that
control the formation of an hexagonal lattice and analyze the role of gravity for curved
interfaces.
\end{abstract}

\keywords{liquid crystals, defects, dynamics, colloidal particles}
\pacs{61.30.Jf,61.30.Hn,64.75.Xc}

\maketitle

Colloidal particles confined at liquid interfaces display rich two-dimensional (2D) phase
properties \cite{collinter1,collinter2}. The spontaneous formation of ordered structures such as
microcrystals has been mainly studied in simple fluids \cite{autoorga1,autoorga2,autoorga3} where
the self-arrangement is controlled  by direct colloidal interactions (electrostatic
\cite{Furst,DANOV2010}, magnetic \cite{collinter1}...) and possible capillary effects. The latter
might come from the anisotropic shape \cite{Oettelellips} or the roughness of the particles
\cite{Krassimir}. It is only recently that an interest
\cite{Nazarenko2004,Nazarenko2007,abbott2008,abbott2010} has developed in the behavior of particles
trapped at an ordered fluid interface. In bulk liquid crystals (LC), additional long-range
interactions between particles are present because of the partial order and elasticity. Colloidal
suspensions \cite{POULIN,POUL97} in a nematic matrix are thus qualitatively different from their
isotropic analogues. They display rich self-ordering phenomena involving particles and topological
defects. At LC interfaces, complex ordered structures were also observed in several cases: glycerin
droplets \cite{Nazarenko2004,Nazarenko2007} or solid beads \cite{Yam2010} at nematic/air interfaces
or microparticules at nematic/water interface \cite{abbott2008,abbott2010}. All these systems
display 2D hexagonal crystals that were ascribed to the competition between a repulsion due to the
bulk liquid crystal elasticity and a capillary attraction resulting from the interface distortions
caused by the ``nematic elastic pressure''. This new type of capillary interaction is however
thoroughly discussed in two recent works \cite{Ottel2009,Pergamenshchik} and its role is not
clearly established. To clarify the respective role of the elastic and capillary force
 a direct force measurement between trapped particles
 coupled with a careful control of  the LC anchoring on the beads as
 well as of the flatness of interface would be suitable.

 In this work, we present a simple technique for trapping
colloids at the flat interface of an
 aligned thin layer of nematic liquid crystal. By controlling the beads density, the interface curvature and the
 LC anchoring, we were then able to establish their respective role in the formation of the colloidal structures.
 A direct measurement of the pairwise interaction has been obtained with optical tweezers, which allowed us to discuss
the respective roles played by LC elasticity and capillarity.
\begin{figure} [ht] \centering
\includegraphics[width=0.43\textwidth]{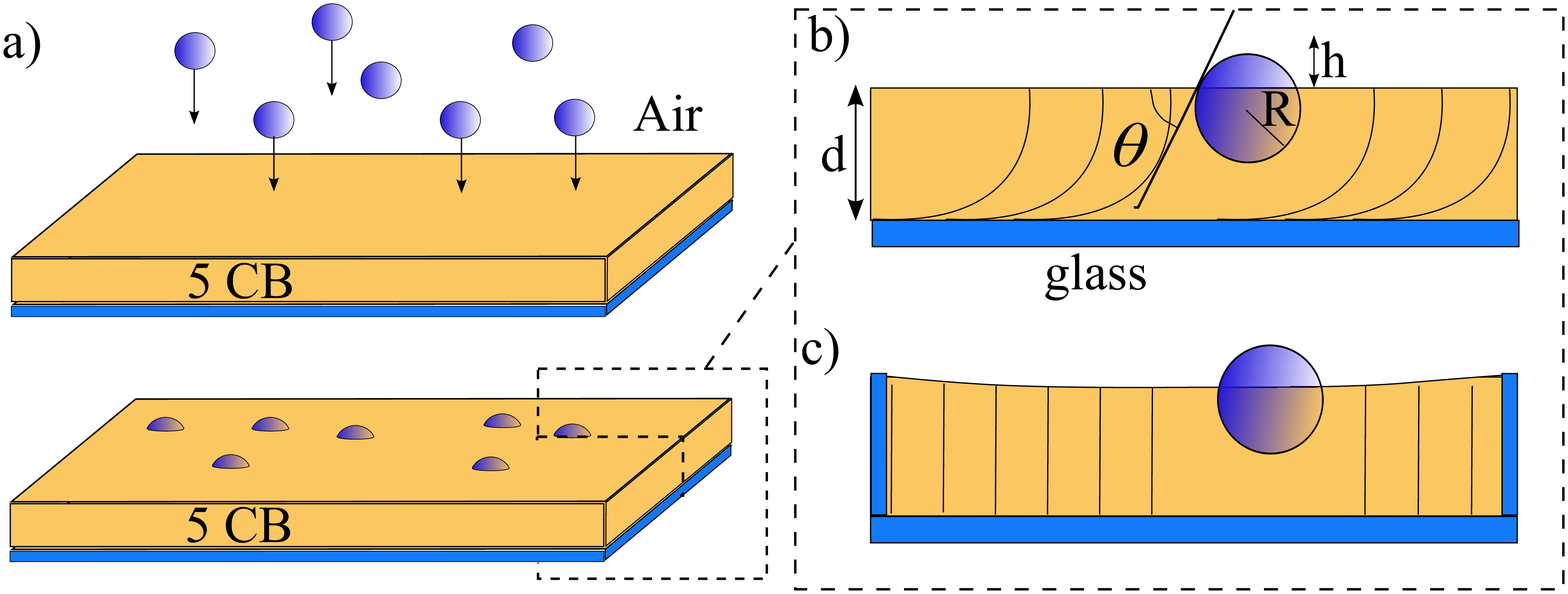}
\caption{a) Deposition of colloids at the air liquid crystal interface. The LC texture is either
hybrid due to the
 strong planar anchoring on polymide and homeotropic at the air (1-b) or fully homeotropic on silanized glass (1-c).}
\label{fig1}
\end{figure}

The studied systems are obtained by trapping solid spheres at the interface between air and a
nematic LC at 21$\degree$C. Aggregates of dry silica beads
 (radius  $R=1.96\mu$m from Bangslabs) are ``exploded''
 by an air pulse in a box. The individual spheres then gently settle on a liquid crystal slab (fig.~\!\ref{fig1}) which
  avoids the presence of colloids in bulk.
  The LC layer (thickness in the range 10-100$\mu$m) is obtained by spin coating
  4-pentyl-4'-cyanobiphenyl (5CB from Synthon)
 on a glass slide treated with polyimide (EHC Japan) that ensures a strong planar anchoring. The layer exhibits
 a hybrid texture due to the strong homeotropic anchoring at the air interface (fig.~\!\ref{fig1}-b).
 Homogeneous homeotropic layers have also been studied by using a silane treatment on glass
with N,N-dimethyl-N-octadecyl-3amininopropyl trimethoxysilyl chloride (DMOAP from Aldrich)
 but the poor wettability of 5CB on silanized surfaces requires using a surrounding glass wall
  as shown in Fig.~\!\ref{fig1}-c. Note finally that the beads -initially
dispersed in water- were covered with a monolayer of DMOAP following Ref. \cite{SKAR}, which
ensures a strong homeotropic surface anchoring on 5CB. They were then dried at $T=110^\circ$C
before use \cite{quality}.

\begin{figure*} [t]
\centering
\includegraphics[width=0.8\textwidth]{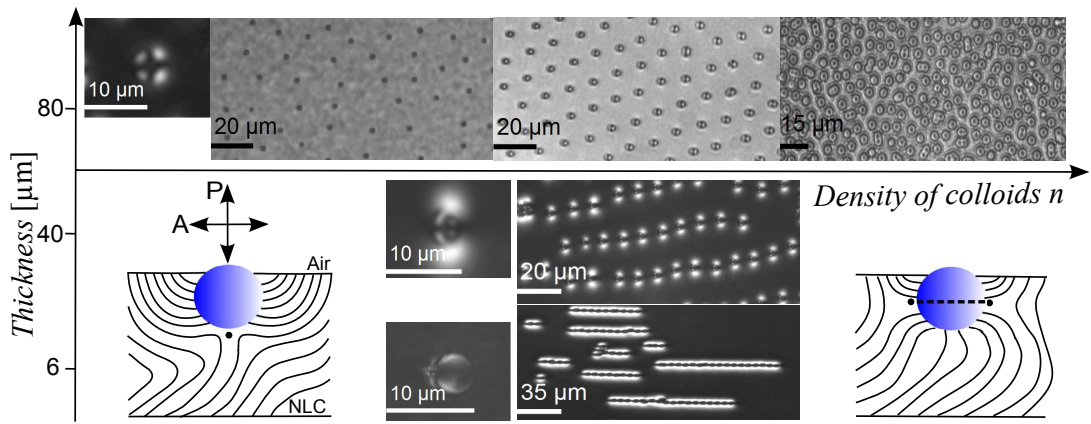}
\caption{Patterns formed by $4\mu$m diameter colloids trapped at the air/NLC interface when a
strong planar anchoring is imposed by the lower substrate. In thin samples, chains form along the
direction of alignment. They are not observed at large thicknesses ($d>40\mu$m) where the patterns
strongly depend on the colloids density, going from a liquid to a loose crystal and an amorphous
condensed state. Sketches illustrate two main possible nematic textures around the beads. }
 \label{fig2}
\end{figure*}

 The colloids/LC systems were
 observed in transmission mode under a
 polarizing microscope (LEICA DM 2500 P) equipped with an INSTEC hot stage (temperature regulated at 0.1$^\circ$C)
 and a SONY 1024x768 digital camera. Birefringence measurements with a Berek compensator were used to determine
 the thickness of the thinnest hybrid films. The microscope also allows an accurate characterization of interfaces
 in reflection mode by Vertical Scanning and Phase Shift Interferometries \cite{CABER,BHUS} thanks to a Mirau
 objective (x20) mounted on a Nano-F (MCL) nanopositionner focusing element. We also used
  optical tweezers based on a LEICA DMI 3000 B inverted microscope equipped with a x100 (NA 1.4) oil immersion
 objective, a 1064nm laser (YLM – 5W from IPG Photonics) and a piezoelectric XY stage (MCL). Silica beads cannot
  be directly trapped because of the inappropriate index contrast in 5CB but can nevertheless be
  manipulated with the 'ghost' effects due to the alignment \cite{GHOST}. Tracking procedures
   (St Andrews Tracker \cite{Labview}) were used to determine accurately the beads position.

We first checked with vertical scanning interferometry that the beads were actually trapped at the
interface. The top of a sphere is easily located with a Mirau objective and the height contrast $h$
(see Fig.\ref{fig1}) with the surrounding fluid additionally gives the contact angle of the beads
at the air interface $\theta =\arccos(1-h/R) =31 (\pm 2)^\circ$. The surrounding fluid is flat
without detectable localized deformation (with a typical vertical resolution of a few nanometers).
After the sample preparation, beads begin to organize into larger clusters, depending on their area
density, the anchoring conditions and the LC layer thickness. The planar case is summarized in
Fig.~\!\ref{fig2}. Between crossed polarizers, a point defect close to the bead is always observed
in thin layers($d < 30 \mu$m). It is reminiscent of the hyperbolic defect that forms around beads
of micrometer size in planar cells \cite{POUL97}. This defect disappears at large thickness, where
the birefringence pattern looks more radial (Maltese Cross in top left). In thin layers, the beads
spontaneously form linear chains parallel to the easy axis. Individual colloids are
 then attracted by those chains which grow and finally collect all surrounding particles.
 When the thickness increases, the chains are much less defined and are no more observed typically
above 40$\mu$m.
 At larger thicknesses, the patterns are very sensitive to the area density of deposited colloids.
 This evolution is shown in the top pictures. At a low colloidal density, a stable liquid behaviour is observed.
  Increasing the density typically above 1000 colloids.mm$^{-2}$,
   crystalline hexagonal domains appear and  form a single crystal in a few
hours. If the density is higher (above 10000 colloids.mm$^{-2}$), an increasing number of amorphous
2D aggregates (top right picture) are observed in coexistence with the crystalline structure. When
the anchoring on the lower substrate is homeotropic the same patterns and density thresholds are
observed indicating that the colloids interactions are very similar in a homeotropic slab and in
large hybrid layers. These observations are somewhat reminiscent of the hexagonal lattices formed
by glycerin droplets at the air-liquid crystal interface \cite{Nazarenko2004,Nazarenko2007}. In
that case, the liquid structure at low density and the amorphous condensed state at large one are
however absent. Two main differences might explain these discrepancies. First the particles we used
are solid silica spheres and are not deformed at the interface. Second, the anchoring on DMOAP is
strongly homeotropic (the anchoring energy is {$W\approx$10$^{-2}$J.m$^{-2}$  \cite{SKAR}}) whereas
it is planar degenerated on glycerin. The detailed nematic texture still has to be deciphered but
two suggestions are sketched  in  Fig. \ref{fig2}. They are based on observations of beads with
homeotropic anchorings dispersed inside nematic planar cells which show the presence of either a
hyperbolic hedgehog point defect or a Saturn ring \cite{POUL97}.
\begin{figure} [ht] \centering
\includegraphics[width=0.4\textwidth]{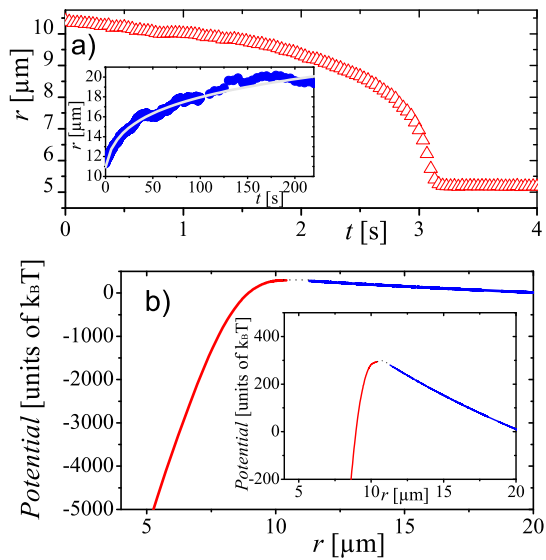}
\caption{Time dependence of the separation distance between two isolated particles released at 10.5
$\mu$m and at 11$\mu$m (Inset). (b) Corresponding pair potential (arbitrarily fixed at $k_BT$ at
r=20$\mu$m) derived from several beads trajectories. Inset: zoom around the unstable equilibrium
distance $r_a$. } \label{fig3}
\end{figure}

 As said above, the nature of the interactions between colloids trapped at a nematic interface is
still debated.  The hexagonal lattices of glycerin droplets \cite{Nazarenko2004} or microparticles
\cite{abbott2010} have been explained by the existence of an equilibrium inter-beads distance. The
latter results from the nematic elastic repulsion competing with a capillary attraction arising
from the nematic pressure on each beads. In a recent paper, Oettel however \cite{Ottel2009} shows
that the weak interface deformation cannot account for the observed effects. The possible role of
many-body interactions in stabilizing the structures has also been discussed recently in
Ref.~\cite{Pergamenshchik}. In our case, the nematic elasticity is clearly  a key ingredient, since
the hexagonal patterns disappear through the nematic-isotropic phase transition. An equilibrium
distance between two particles is however hardly compatible with the various observed structures
and we focused on the bead-bead pair potential. With optical tweezers, two isolated beads are
approached at an initial distance $r_\circ$ and are tracked after trap release. We show in
Fig.\ref{fig3}(a) a typical evolution of the separation distance $r$. As long as $r_\circ$ is
larger than $r_a = 10.5\pm0.5 \mu$m, the beads move away from each other. They irreversibly
aggregate for $r_\circ<r_a$ with a final separation of 1$\mu$m. Averaged over several trajectories,
the interaction force $f_p$ can be obtained from the Stokes law \cite{STOK}, and the pair potential
energy $E_p$  by its integration \cite{Musevic2007}. The latter is given in Fig.\ref{fig3}(b). For
beads with {\em homeotropic} anchoring, our observations therefore prove that an {\em unstable}
equilibrium distance $r_a$ separates a region of attraction at short distances, and of repulsion at
larger ones. The value of $r_a$ is roughly constant with the thickness (above 40 $\mu$m) and the
pair orientation, indicating that the lower planar anchoring is screened by the homeotropic
anchoring on air. Such a pair energy profile differs from the one between two glycerin droplets
expected in Ref.~\cite{Nazarenko2004}. We first checked if the repulsive part between two trapped
beads was compatible with a ``pure bulk'' elastic interaction in a large-distance multipolar
development. Whatever the exact nematic texture, the homeotropic anchoring at the air forbids a
dipolar distortion \cite{Ottel2009} around a bead (for textures with cylindrical symmetry).
Dimensional analysis yields the following pair potential for the non-zero quadrupole moment
\cite{LUB}:
\begin{equation} \centering
 E_p=\frac{36\pi K \beta^2 R^6}{r^5} \label{qpole},
\end{equation}
where $K\sim10^{-11}$N is the 5CB elastic modulus in the one constant approximation and  $\beta$ a
 coefficient of order unity. The repulsive trajectory is then given by the competition
between the drag force  $f_v=-\gamma v=-\gamma \dot{r}/2$ and the driving force $f_p=-\partial E_p
/\partial r$:
\begin{equation} \centering
r(t)=(2520  \pi K \beta^2 R^6 t/\gamma+r_\circ ^7)^{1/7}.
\end{equation}
 This expression correctly fits the trajectories (Fig.~\!\ref{fig3}-a) with $\beta = 2.1\pm0.2$.
  To explain the short distance attraction, we first examined the simple
approach of Ref.~\cite{Nazarenko2004} that consists in adding a capillary attraction term. The
corresponding logarithmic dependence is however overwhelmed by the algebraic elastic dependence at
short distance. A more refined approach derived from Ref.~\cite{Ottel2009} yields the same
conclusion: an additional capillary attraction is unlikely responsible for the observed attraction.
A strong reorganization of the director field  could however explain it, as already observed in the
short distances binding observed for homeotropically-treated beads in bulk nematic films
\cite{Zumer2006} and also supported by the similitude of the pair potential profile with the one
theoretically computed for infinite parallel cylinders located at the nematic/isotropic interface
in Ref.~\cite{N/I} where such a reorganization clearly appears at short distances.
\begin{figure} [ht] \centering
\includegraphics[width=0.42\textwidth]{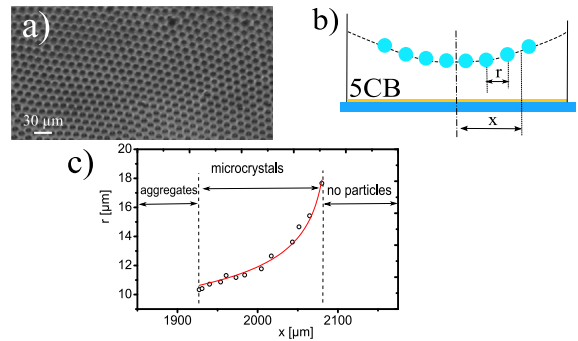}
\caption{(a) Optical micrograph of hexagonal crystals at a curved nematic/air interface. (b)
Geometry of the experiment (c) Measured separation distance $r$ between colloids as a function of
its distance $x$ from center.
 } \label{fig4}
\end{figure}

The  pairwise potential qualitatively explains the structures formed by the trapped beads. At low
densities, the mean distance $r_m$ between two beads is much larger than $r_a$ and  a liquid is
observed. The hexagonal patterns spontaneously form for intermediate densities where $r_a<r_m$) due
to the strong inter-beads repulsion. The lattice period is however limited by $r_a$ and for large
enough densities ($r_m<r_a$) some particles spontaneously and irreversibly aggregate. Differently
from the glycerin droplets case, the
 period is therefore not univocally defined but changes with area density. To probe if
 many-body effects strongly influence this scenario,
  a known additional interaction was applied on the crystals.
 We used slightly convex air-NLC interfaces resulting from a non perpendicular contact angle on a
cylindrical clean glass tube (see Fig.~\!\ref{fig1}-c). After the deposition (at a weak density),
the beads converge to the center. After a few days they form stable large crystals with a period
depending on the distance $x$ from the center. When more beads are added at the periphery the
system reaches a new equilibrium with a decreasing mesh size until aggregates start to form in the
center. Figure \ref{fig4}-c shows the lattice period as a function of $x$. A single amorphous
aggregate is observed at the center. Further, the crystal appears with a period close to $r_a$ that
rapidly increases up to the crystal end. Such a variation is due to the gravity forces projected
onto the interface that oppose the beads' repulsion. The resulting 2D pressure $\Pi=\sqrt{3}f_p/r$
is derived from the inter-beads force in an hexagonal lattice of density $n=2/\sqrt{3}r^2$. The
balance of the forces yields:
 \begin{equation} \centering
 \Delta mg \sin(\alpha x)n(x)=-\frac{\partial \Pi(x)}{\partial x},
\end{equation}
where $\Delta m$ is the effective mass of a buoyant trapped bead and $\alpha x \ll 1$  the
interface slope. Using (\ref{qpole}) as the pair potential, we obtain the form
 \begin{equation} \centering
 r=\left(A-Bx^2\right)^{-1/5},
\end{equation}
where $A$ is an integration constant and $B=\Delta mg \alpha/756 \pi K \beta^2 R^6$. This
expression fits the data  with $B\approx1.1 \times 10^{31}$ m$^{-7}$ (Fig.\ref{fig4}-c) while we
measured $\alpha =92.2$m$^{-1}$. Although this result yields a slightly smaller value $\beta
\approx 1.4$ when taking for densities $\rho_{beads}\approx 2$ g.cm$^{-3}$ (provided by Bangslabs)
and $\rho_{5CB}\approx$ 1 g.cm$^{-3}$, we can conclude that the pairwise potential is enough to
describe the observed patterns at least in a first approximation.

In conclusion, we have exhaustively described the behavior of microparticles with a homeotropic
anchoring at a nematic/air interface and reported on the first direct measurements of the
corresponding pairwise interaction. The latter is found attractive at short distances and repulsive
at large ones. The long-range repulsive part is compatible with an elastic quadrupolar interaction.
It satisfactory accounts for the hexagonal crystals we observed under simple or gravitational
confinement. This same repulsive interaction could also be at the origin of the recently observed
 crystals at  LC/water interface \cite{abbott2010}.
The origin of the attractive part is still an open question. The measured flatness of the interface
and the expected spatial dependency of the capillary interaction suggest it also has an elastic
origin. One possible mechanism of this all-elastic interaction from repulsion to attraction could
be related to a defect transformation with consequent colloidal binding. We are confident that this
work will stimulate further experimental and theoretical studies to elucidate these fascinating
interfacial phenomena.

 \acknowledgements
This work was supported in part by  French ANR grant BLAN07-1\_183526 ``Surfoids''. The authors
also thank M. Abkarian for fruitful discussions.
 

\vspace{-0.3cm}
\begin{thebibliography}{99}
\bibitem{collinter1} U. Gasser {\em et al.}, ChemPhysChem., \textbf{11}, 963 (2010).
\bibitem{collinter2} B. P. Binks, Phys. Chem. Chem. Phys. , \textbf{9}, 6298 (2007).
\bibitem{autoorga1} P. Pierensky, Phys. Rev. Lett., \textbf{45}, 569 (1980).
\bibitem{autoorga2} T. Terao and T. Nkayama, Phys. Rev. E, \textbf{60}, 7157 (1999).
\bibitem{autoorga3} A. D. Dinsmore {\em et al.}, Science, \textbf{298}, 1006 (2002).
\bibitem{DANOV2010} K. D. Danov, and P. A. Kralchevsky, J. Colloid Interf. Sci.  \textbf{345}, 505
(2010).
\bibitem{Furst} B. J. Park {\em et al.}, Langmuir, \textbf{24}, 1686 (2008).
\bibitem{Oettelellips} E. Nouruzifar, and M. Oettel, Phys. Rev. E, \textbf{79}, 051401 (2009).
\bibitem{Krassimir} K. D. Danov, and P. A. Kralchevsky, Adv. Colloid Interface Sci. \textbf{154}, 91 (2010).
\bibitem{Nazarenko2004} I. I. Smalyukh {\em et al.} , Phys. Rev. Lett. \textbf{93}, 117801 (2004).
\bibitem{Nazarenko2007} A. B. Nych {\em et al.}, Phys. Rev. Lett. \textbf{98}, 057801 (2008).
\bibitem{abbott2008} I. H. Lin {\em et al.}, J. Phys. Chem. B \textbf{112},16552 (2008).
\bibitem{abbott2010} G. M. Koeing {\em et al.},Proc. Natl. Acad. Sci. USA \textbf{107},3998 (2010).
\bibitem{POUL97}P. Poulin and D. A. Weitz, Phys. Rev. E \textbf{57}, 626 (1998).
\bibitem{POULIN} P.Poulin {\em et al.}, Nature \textbf{275}, 1770 (1997).
\bibitem{Yam2010} T. Yamamoto, and M. Yoshida, Appl. Phys. Express \textbf{2}, 101501 (2009).
\bibitem{Ottel2009} M. Oettel {\em et al.}, Eur. Phys. J. E \textbf{28}, 99 (2009).
\bibitem{Pergamenshchik} V. M. Pergamenshchik, Phys. Rev. E \textbf{79}, 011407 (2009).
\bibitem{SKAR} M. \v{S}karabot {\em et al.}, Phys. Rev. E \textbf{77}, 031705 (2008).
\bibitem{quality} The anchoring was checked by dispersing some beads in a planar
5CB cell where colloidal chains and distinctive birefringence spontaneously patterns form
\cite{SKAR,POUL97}.
\bibitem{CABER} P. J. Caber, Appl. Opt. \textbf{32}, 3438 (1993).
\bibitem{BHUS} B. Bhushan, J. C. Wyant and C. Koliopoulos, Appl. Opt. \textbf{24}, 1489 (1985).
\bibitem{GHOST} I. Mu\v{s}evic {\em et al.}, Phys.Rev.Lett. \textbf{93}, 187801 (2004).
\bibitem{Labview} C. Lopez-Mariscal {\em et al.}, Opt. Express \textbf{14},
4182 (2006).
\bibitem{STOK}The Stokes's law $f=\gamma v$
 acting at the velocity $v$ is determined
from the diffusion coefficient $D=k_{B}T/\gamma$ obtained from the mean squared displacement of an
isolated bead.
\bibitem{Musevic2007} M. \v{S}karabot {\em et al.}, Phys. Rev. E \textbf{76}, 051406 (2007).
\bibitem{LUB} T. C. Lubensky {\em et al.}, Phys. Rev. E \textbf{57}, 610 (1998).
\bibitem{Zumer2006} I. Mu\v{s}evic {\em et al.}, Science \textbf{313}, 954 (2006).
\bibitem{N/I} D. Andrienko,M. Tasinkevych, and S. Dietrich, Europhys. Lett., \textbf{70}, 95
(2005).
\end{thebibliography}
\end{document}